# Exploring the Societal and Economic Impacts of Artificial Intelligence: A Scenario Generation Methodology

Carlos J. Costa
Advance/ISEG (Lisbon School of Economics & Management), Universidade de Lisboa
Lisbon, Portugal
cjcosta@iseg.ulisboa.pt

Joao Tiago Aparicio
Instituto Superior Técnico, Universidade de Lisboa
Lisbon, Portugal;
joao.aparicio@tecnico.ulisboa.pt

## ABSTRACT
This paper explores artificial intelligence's potential societal and economic impacts (AI) through generating scenarios that assess how AI may influence various sectors. We categorize and analyze key factors affecting AI's integration and adoption by applying an Impact-Uncertainty Matrix. A proposed methodology involves querying academic databases, identifying emerging trends and topics, and categorizing these into an impact uncertainty framework. The paper identifies critical areas where AI may bring significant change and outlines potential future scenarios based on these insights. This research aims to inform policymakers, industry leaders, and researchers on the strategic planning required to address the challenges and opportunities AI presents.

•**Social and professional topics** → **Professional topics** → **Computing and business** → **Economic impact**

•**Social and professional topics** → **Professional topics** → **Computing and business** → **Employment issues**

•**Social and professional topics** → **Computing / technology policy** → **Government technology policy** → **Governmental regulations**

## 1. INTRODUCTION
The increasing integration of artificial intelligence (AI) across various sectors raises important questions about its societal and economic implications. AI has the potential to revolutionize industries by enhancing efficiency, generating new economic opportunities, and reshaping societal structures[1]. However, its rapid development also introduces significant uncertainties regarding its long-term trajectory. Understanding these impacts is essential for policymakers, businesses, and society to navigate AI's opportunities and challenges [2].

This study seeks to answer the following central question: How will AI shape society and the economy in the coming years? To address this, the research explores and categorizes potential scenarios for AI's societal and economic impact. The key objectives include:

- Identifying the primary areas where AI is expected to drive transformative changes.
- Assessing the opportunities and risks associated with AI's integration into different sectors.
- Developing structured scenarios that outline possible future developments in AI-driven economies and societies.

This study employs a robust methodological framework combining academic literature analysis and scenario generation to achieve these objectives. A systematic review of existing research will provide a foundation for understanding current trends and debates surrounding AI's impact. Scenario generation techniques will then be applied to construct plausible futures, considering diverse economic, technological, and societal variables.

This paper is structured as follows: Section 2 reviews the existing literature on AI's societal and economic impacts. Section 3 presents the research methodology, detailing the scenario-building approach. Section 4 outlines the key findings, including identified scenarios and their implications. Finally, Section 5 discusses the broader implications of these findings and provides recommendations for policymakers and industry leaders.

## 2. LITERATURE REVIEW
The growing influence of artificial intelligence (AI) across industries has stimulated extensive research into its economic and societal effects. Scholars have examined AI's role in transforming key sectors such as healthcare, education, finance, and energy, as well as its broader implications for labor markets, economic growth, and governance [1]. Ethical considerations, regulatory challenges, and public trust in AI systems remain central themes in ongoing debates[2] [3].

A significant body of research has focused on forecasting AI's long-term impacts. Studies highlight AI's potential to enhance human capabilities, such as improving diagnostic accuracy in healthcare and enabling personalized learning in education.[4] However, concerns about job displacement, economic inequality, and the governance of AI systems have led to increased interest in predictive methodologies that anticipate AI-driven disruptions. In particular, scenario-generation approaches have been employed to explore different pathways AI adoption might take, helping policymakers and businesses prepare for future developments.[5]. As AI adoption accelerates, researchers are increasingly interested in its structural effects on economies and societies.

Regarding the economic impacts of AI, the following dimensions are being specially analyzed by researchers: Economic Growth[6], Capital Accumulation [7], and Employment Patterns[8].

AI has a significant positive effect on economic growth, particularly when considering the population's external system. However, the impact varies by the level of AI development, with middle levels contributing significantly to growth while high levels show no significant contribution [6].

AI is used as a capitalist tool that aids corporate monopolies and creates alienating development, impacting economic structures and governance [7].

AI is predicted to cause significant shifts in employment, with a high probability of AI-induced unemployment reaching 40-50% by 2040. This shift necessitates urgent policy responses to manage economic inequalities and societal fragmentation [8].

Concerning the Societal Impacts of AI, researchers are considering Public Services [9], Social Structures[10] and Sustainable Development [11]

A public value perspective methodology can assess AI's impact on public services, identifying opportunities, threats, enablers, and barriers [9].

AI technologies impact social structures by optimizing production processes and developing human resources, but they also pose risks to economic resilience and social security [10].

AI affects all three pillars of sustainable development—economic, social, and environmental. It optimizes resource use and increases energy efficiency but also disrupts labor markets by discrediting and potentially eliminating hundreds of professions [11]

Among the approaches used to predict, the researchers have suggested several approaches [5] [4]. Among them, scenario generation methodology is used. For example, Delphi Process and Probabilistic Modelling [8], or Policy Scenario Methodology [12].

Utilizing a Delphi process coupled with probabilistic modeling can construct detailed scenarios that reveal the cascading effects of AI across economic, societal, and security domains [8].

This approach involves constructing likely future scenarios based on current trajectories to articulate issues and develop foresight analysis. It helps identify key facilitators and inhibitors of change and the societal impacts of AI implementation [12].

Among the key considerations for scenario generation are relevant regulatory challenges [4] [13] and ethical [13] and social justice [14].

The rapid evolution of AI technologies often outpaces existing governance frameworks, exacerbating economic inequalities and societal fragmentation.[15] Proactive governance and robust regulatory frameworks are essential to harness AI's benefits without compromising global stability [8] [16].

Ensuring that AI advances are compatible with democratic values and social justice requires interdisciplinary collaboration among technical experts, ethicists, lawyers, and social scientists. Appropriate regulations and ethical guidelines are crucial for achieving a just society [14].

## 3. METHODOLOGY

This research aims to identify and analyze scenarios where AI could significantly impact society and the economy. We propose a scenario-generation methodology that involves querying academic databases, analyzing trends, and constructing an Impact-Uncertainty Matrix to guide strategic decision-making. The methodology involves the following steps:

1. Query selection
2. Use of query in Scopus
3. Topic Identification and Categorization
4. Constructing the Impact-Uncertainty Matrix
5. Scenario Generation
6. Scenaeos quantification

This research started by creating a query from Scopus academic databases to identify key scholarly works that explore AI's societal and economic implications. The search query focuses on AI, ethics, policy, societal impacts, and future trends. This query is the following:

```
TITLE-ABS-KEY ( ("artificial intelligence" OR
"AI") AND ("ethics" OR "societal impact" OR
"policy" OR "regulation" OR "economy") AND
("predictions" OR "trends" OR "future
directions" OR "forecasts") )
```

A comprehensive dataset of references that focus on AI's societal and economic impact was gathered from the search query. These articles provide insights into challenges, trends, and predictions for AI's future.

The next step involves categorizing the identified articles into broad themes or topics based on their content. These topics represent the areas where AI is expected to have the most influence.

An Impact-Uncertainty Matrix is created to evaluate each identified topic based on its potential societal and economic impact and uncertainty. This matrix helps prioritize topics that require strategic attention.

Based on the Impact-Uncertainty Matrix, we generate possible future scenarios considering both high-impact and uncertainty factors. These scenarios provide a framework for understanding how different factors may evolve and inform policy and decision-making.

## 4. RESULTS

## 4.1 Defining Dimensions

After collecting the dataset, topic modeling was performed. First, the text was vectorized using Tf-idf Vectorizer[17]. Then, the Latent Dirichlet Allocation [2] was used, and both used sklearn implementation[18].[19]

The results obtained are the following:

```
== Identified Topics ===
Topic 1: ai research digital data intelligence
artificial development technology technologies
education --> Social, Technological
Topic 2: gene protein expression drug cell cancer
molecular rna breast biology --> Unclassified
Topic 3: learning model prediction data models
machine neural deep network forecasting -->
Technological
Topic 4: energy power carbon renewable emissions
electric solar grid wind electricity -->
Environmental
Topic 5: financial stock market price insurance
banking credit fintech finance markets --> Economic
Topic 6: health care medical clinical disease
healthcare patient covid medicine patients -->
Social
```

The result may also be represented in the following chart.

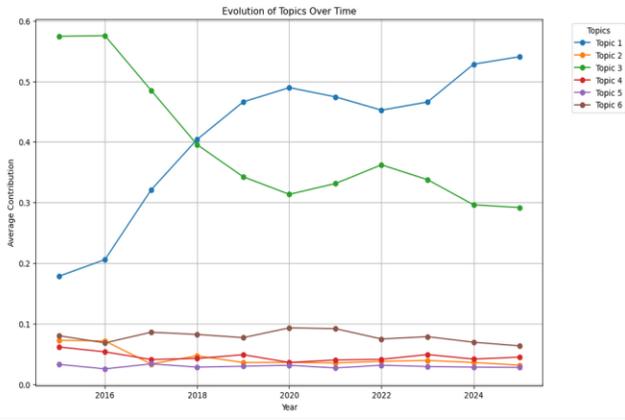

**Figure 1.** Evolution of relative frequency of topic references.

Using the query outlined in the methodology, we identified six key topics where AI could have significant impacts:

- Topic 1 - "AI and Digital Education
- Topic 2 - "Molecular Medicine and Oncology
- Topic 3 - "Machine Learning and Predictive Analytics"
- Topic 4 - "Renewable Energy and Sustainability"
- Topic 5 - "Financial Markets and Fintech"
- Topic 6 - "Healthcare Systems and Public Health".

Topic 1, "AI and Digital Education," focuses on artificial intelligence research, digital data, and the integration of technology in education. Topic 2 corresponds to "Molecular Medicine and Oncology". It centers on gene expression, drug development, cancer research, and molecular biology. Topic 3, "Machine Learning and Predictive Analytics," covers learning models, data analytics, neural networks, and forecasting technologies. Topic 4, "Renewable Energy and Sustainability," addresses energy production, renewable resources, and environmental sustainability. 5. topic 5 corresponds to "Financial Markets and Fintech," which explores financial markets, banking, insurance, and financial technology innovations. Topic 6 is "Healthcare Systems and Public Health". It focuses on healthcare services, clinical practices, disease management, and public health issues, including COVID-19.

An Impact-Uncertainty Matrix is a strategic planning tool that helps prioritize factors based on their potential impact and the uncertainty surrounding them. Here is how you can structure the matrix for the identified topics.

Impact refers to the potential influence of the factor on society, economy, or environment. High-impact factors significantly shape future scenarios. Uncertainty refers to the level of unpredictability associated with the factor. High uncertainty factors are difficult to forecast and require flexible strategies.

Factors with high impact and high uncertainty are critical and require proactive and adaptive strategies. Factors with high impact but moderate uncertainty are important and need strategic attention but may be more manageable. This matrix helps identify which factors need the most attention in scenario planning, allowing for a more focused and effective strategic approach.

Based on the Impact-Uncertainty Matrix, the three most critical dimensions are those with high impact and high uncertainty. These dimensions are essential for scenario planning because they have the potential to influence future outcomes significantly and require strategic attention due to their unpredictability.

**Table 1. Impact-Uncertainty Matrix**

| Factor/ Topic | Impact Level | Uncertainty Level | Strategic Relevance | Implications & Challenges | Emerging Strategies |
|---|---|---|---|---|---|
| AI & Digital Education | High | High | Critical | Transforms learning with personalized and adaptive methods; risks include widening the digital divide and ethical concerns. | Invest in robust digital infrastructure, develop ethical AI guidelines, and promote digital inclusion. |
| Molecular Medicine & Oncology | High | Medium | Important | Promises breakthroughs in targeted therapies and diagnostics yet faces regulatory, ethical, and translational challenges. | Enhance R&D funding; streamline clinical trial and approval processes; ensure ethical oversight. |
| Machine Learning & Predictive Analytics | High | Medium | Important | Drives innovation in multiple sectors; issues of data governance, transparency, and algorithmic bias persist. | Strengthen data transparency and fairness; invest in scalable, secure ML frameworks. |
| Renewable Energy & Sustainability | High | High | Critical | Essential for combating climate change, uncertainty in technological breakthroughs and market adoption remains. | Support clean technology R&D; provide policy incentives; upgrade energy infrastructures. |
| Financial Markets & Fintech | High | High | Critical | Disrupts traditional finance with digital innovations but may increase market volatility and cybersecurity risks. | Implement robust regulatory frameworks; foster innovation in fintech; enhance digital literacy. |
| Healthcare Systems & Public Health | High | Medium | Important | Can revolutionize care delivery and patient management; challenges include data privacy, system integration, and equity. | Prioritize secure and interoperable health data systems; invest in telemedicine; enforce strict privacy protocols. |

The most relevant dimensions are:

1. AI and Digital Education
2. Renewable Energy and Sustainability
3. Financial Markets and Fintech

The next step is to generate potential future scenarios based on the dimensions identified in the Impact-Uncertainty Matrix. These scenarios provide insights into how AI could evolve in various contexts.

## 4.2 Defining Scenarios

The key areas identified as most relevant for scenario planning due to their high impact and high uncertainty. This refined matrix provides a focused framework for exploring how different scenarios might affect these critical dimensions, helping to identify potential challenges and opportunities in future planning.

**Table 2. Future AI Scenarios**

| Scenario | AI & Digital Education | Renewable Energy & Sustainability | Financial Markets & Fintech |
|---|---|---|---|
| Optimistic Future | Rapid, inclusive AI integration leads to personalized, adaptive learning environments. | Accelerated transition to renewables with widespread grid modernization. | Stable and secure digital financial ecosystems bolstered by strong fintech growth. |
| Technological Stagnation | Slow integration with persistent digital divides and outdated pedagogical methods. | Limited breakthroughs with reliance on legacy energy systems. | Heightened market volatility due to sluggish fintech adoption and minimal innovation. |
| Sustainability Focus | Targeted interventions ensuring equitable access and blended learning environments. | High investments in renewable infrastructure driven by strong climate policies. | Stable markets oriented toward green finance, with moderate fintech innovation. |
| Economic Downturn | Budget constraints lead to reduced funding for edtech innovation and digital initiatives. | Minimal investments in renewables result in compromised infrastructure development. | Declining markets and reduced fintech activity elevate exposure to financial risks. |

The following scenarios present different potential futures that consider how each dimension might evolve:

- Optimistic Future
- Technological Stagnation
- Sustainability Focus
- Economic Downturn

The optimistic future assumes rapid technological progress and positive outcomes across dimensions. At the same time, Technological stagnation assumes limited progress and challenges in innovation. The sustainability focus scenario prioritizes environmental sustainability. The economic downturn considers the impact of economic decline on each dimension.

## 4.3 Quantify the scenarios

The purpose is to model the evolution of three key indices over time: Economic Growth (E), Social Well-being (S), and Technology Advancement (T). Two primary factors influence those indicators. AI and Digital Education Level (A) and Renewable Energy and Sustainability Level (R) are those factors. The mathematical formulations underlying this simulation incorporate elements from well-established growth models, including the Logistic Growth Model and the Gompertz Function, to capture realistic growth dynamics.

In this model, the following Variables and Parameters may be identified:

- **A** - AI and Digital Education Level (ranging from 0 to 1).
- **R** - Renewable Energy and Sustainability Level (ranging from 0 to 1).
- **E₀, S₀, T₀** - Initial values for Economic Growth, Social Well-being, and Technology Advancement, respectively.
- **α, β, γ** - Parameters representing the contributions of A, R, and their interaction to Economic Growth.
- **δ, ε, ζ** - Parameters representing the contributions of A, R, and their interaction to Social Well-being.
- **η, θ** - Parameters representing the contributions of A and the interaction of A and R to Technology Advancement.
- **k_e, k_s, k_t** - Growth rate constants for E, S, and T, respectively.
- **Σ** - Standard deviation for stochastic variation (noise).

The evolution of Economic Growth is modeled using a Logistic Growth Function, which describes how growth accelerates rapidly at first and then slows as it approaches a saturation point. The equation is:

$$E_t = \frac{E_0 + \alpha A + \beta R}{1 + e^{-(k_E(t-t_0))}} + \epsilon_E$$

Where:

- E₀ is the initial economic growth value.
- αA represents the contribution of AI and Digital Education.
- βR represents the contribution of Renewable Energy and Sustainability.
- k_e is the growth rate constant.
- t₀ is the inflection point where the growth rate changes
- εE~N(0,σ) (Noise) introduces stochastic variation to simulate real-world unpredictability.

The logistic function is commonly used to model population growth and other phenomena where growth is self-limiting due to resource constraints.

Similar to Economic Growth, Social Well-being is modeled using a Logistic Growth Function:

$$S_t = \frac{S_0 + \delta A + \epsilon R}{1 + e^{-(k_S t - 5)}} + \epsilon_S$$

Where:

- S₀ is the initial social well-being value.
- δA represents the contribution of AI and Digital Education.
- εR represents the contribution of Renewable Energy and Sustainability.
- k_s is the growth rate constant.
- t₀ is the inflection point.
- Noise adds stochastic variation.

Technology Advancement is modeled using the Gompertz Function, which is characterized by the slowest growth at the start and end of a time period, with the fastest growth occurring in between. The equation is:

$$T(t) = (T_0 + \eta A + \theta AR)e^{-e^{-k_T(t-t_0)}} + \epsilon_T$$

Where:
- T₀ is the initial technology advancement value.
- ηA represents the contribution of AI and Digital Education.
- θA represents the synergistic effect of AI and Sustainability.
- $k_t$ is the growth rate constant.
- $t_0$ is the inflection point.
- Noise introduces stochastic variation.

The Gompertz function is widely used to describe growth processes where the rate of growth decreases exponentially over time, such as in tumor growth modeling.

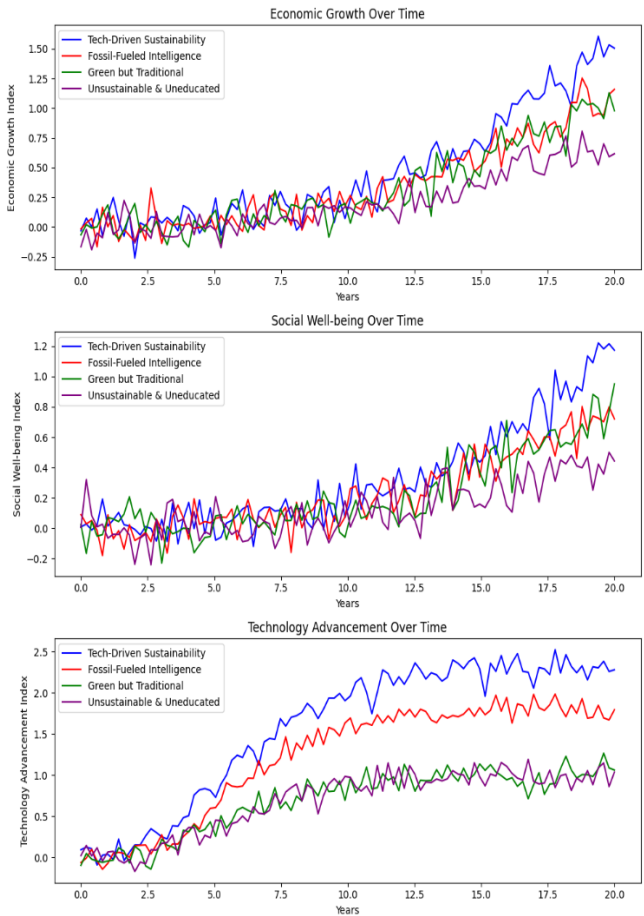

**Figure 2. Evolution of economic growth over time, Social well-being over time, and Technological advancement over time.**

The logistic growth functions are used for Economic Growth and Social Well-being. These functions reflect scenarios where growth accelerates initially but slows as it approaches a maximum limit, representing saturation effects due to limited resources or other constraints.

The Gompertz function is applied to technology advancement. This function captures the characteristics of rapid technological growth during a middle phase but slows down as it reaches maturity, reflecting challenges in sustaining innovation rates over time.

The inclusion of noise in each equation accounts for randomness and unpredictability in real-world scenarios, providing a more realistic simulation of how these indices might evolve over time.

## 5. DISCUSSION

The Impact-Uncertainty Matrix revealed that AI in education, renewable energy, and fintech are the most critical areas for strategic planning. These fields are marked by high impact and uncertainty, making them key factors to monitor in future AI developments.

The potential for AI to transform education is immense, but challenges such as infrastructure limitations, ethical concerns, and equitable access may impede progress. Proactive investment in these areas is necessary.

AI's ability to optimize energy use and promote sustainability is critical for addressing global climate challenges. However, uncertainty around political and economic factors and the technological readiness of renewable resources requires a flexible strategy.

AI reshapes financial markets and fintech. It brings opportunities and risks. The rapid pace of innovation in these sectors and market volatility necessitates careful monitoring and adaptive policy responses.

One of the relevance of this paper is also the proposal of a scenario analysis incorporating machine learning techniques. This approach still has the possibility of being improved.

## 6. CONCLUSIONS

This paper has outlined a methodology for identifying and analyzing scenarios regarding AI's societal and economic impacts. Using an Impact-Uncertainty Matrix, we have identified key areas where AI may significantly affect society, including education, energy, and finance. By exploring various future scenarios, we provide a strategic framework to help policymakers, businesses, and researchers better prepare for AI's challenges and opportunities. As AI evolves, ongoing research and strategic investment in these critical areas will be essential to ensure positive outcomes and mitigate potential risks.

This paper presented a relevant contribution: the proposal of a scenario analysis approach incorporating machine learning techniques. It also opens the opportunity for future improvement.

## 7. ACKNOWLEDGMENTS

This research was funded by FCT—Fundação para a Ciência e Tecnologia, I.P. (Portugal), under research grant numbers ADVANCE-CSG UIDB/04521/2020